\documentclass[reqno, centertags]{amsart}
\usepackage{amssymb}
\usepackage{graphicx}

\theoremstyle{definition}

\begin{document}

\title{Questions and Physical Reality\\ \emph{- simple philosophical considerations-}}

\author{Michele Caponigro}

\address{Physics Department, University of Camerino, I-62032 Camerino, Italy }

\email{michele.caponigro@unicam.it}

\author{Helen Lynn}

\address{Quantum Philosophy Theories, www.qpt.org.uk }

\email{helen.lynn@qpt.org.uk}

\date{\today}

\begin{abstract}
We argue, through some philosophical considerations, on
(i)dependent or (ii) an independent existence of physical reality underlying quantum states.
According these simple considerations, we conclude that is impossible to have a clear
independent existence of physical reality, we need to search the reasons in the
relationship between our questions (the observer) and the consequent answers (always estimated by the same observer).
Finally, we infer that every theory is affected by our "questions", so we cannot speak about an \textbf{unconditional}
and independent theory underlying physical reality. Plan of the paper, the existence of physical reality underlying quantum states: (i) it
before bit, (ii)it without bit,(iii)it from bit.

\end{abstract}

\maketitle
\section{Assumptions}
We assume that:
\begin{enumerate}
\item $\mid\Psi>$ provide a {\bf complete} description of physical properties.
\item The observer is not an experimental device.
\item The observer is the only ontic element of physical reality.
\item Only the ontic observers compare different quantum states.
\item The theories elaborated by observers could be (i) ontic
(supported by experimental data) or (ii) epistemic.
\end{enumerate}

We will show that under these conditions we have serious difficult to proof an independent existence
of physical reality, only we could to proof his "dependent" existence, an existence after the observer.

\section{It before bit}

We know about the complex relationship between questions and
answers, for instance, when does a statement constitute an answer
to question? How can it be known that the statement is a correct
answer to a question? The answer to these questions depends on the
\textbf{degree of certainty required}, a formal definition of the
problem domain, language, and theory is required to produce
reliable results. In this paper we shall try to argue on the possible
"a priori" or "a posteriori" existence of quantum states, from philosophical point of
view,our objective is to find a criterium to proof a possible choice
utilizing three possibility: first, \textbf{it before bit.} With this
statement we mean the emergence of physical reality independent
from any observers, in this case we assume that \textbf{the observer do not exist.} Can we find, in
this case,the "independent" existence of quantum states? Realistic
approach to quantum mechanics seem to belong to this case, we
disagree, we argue that is not possible to establish any criterium
for their existence, \textbf{we cannot say anything.} Every good
realistic position do not leave from the presence of the observer.
\section{It without bit}
This second approach seem ideal, the observer exist \textbf{but is silent},
a kantian system that include an embarrassing silent observer, is not considered
in our philosophy. Eastern philosophy ,instead, have a long tradition on this subject. In this
paper we do not venture in this field. \textbf{It without bit}, seem to
belong to \textbf{platonic system}, most researches thinks to be silent
and Fapp. We agree with Fapp but they are not silent. They ask many things as observers,
but in their theories the same observer has not any relevance.In any case, like previous analysis,
we cannot proof any clear "a priori" existence of the physical reality, also this approach seem
inadequate.\\
\textbf{Observation:} Some analysis by Zeilinger \cite{Zei} seem quite close this approach, he does not need any observer,
a device is sufficient to deduce the underlying physical reality of quantum states.

\section{It from bit}
Third case, seem we have find the right approach, in this case we have
an active observer, the observer decide the fundamental questions
in order to find the fundamental answers. Seem a clear kantian
position, but our position is different. The emergent physical reality seem
linked with the "questions" of the observers, this conclusion is not so appreciated
from scientific community, it show a negative impression: a subjective
method, but as we seen, from philosophical point of view, \textbf{is the only} possible.
We remember the assumption of ontic observer.
Now, as observer we try to proof the only possible "independent" \textbf{subjective} existence of quantum states.
\section{It from bit: examples of subjective "a priori" existence of quantum states}
Premise:
\begin{enumerate}
\item Quantum states provide a {\bf complete} description of physical properties.
\item 2.1 Fist assumption: We need to postulate the existence of the observer for his description. \\2.2 Second assumption:
every observer is a physical property.
\end{enumerate}

\textbf{Quantum states provide a complete description of physical
reality (observer included).}\\

\textbf{Observation:} For instance, Rovelli's program \cite{Rov1} follow
this approach to quantum mechanics.  We argue that this approach
will be considered the same "subjective" for the inevitable
presence of the observer. According Rovelli instead the approach
is not subjective. The denied observers compare the different quantum states not the physical systems.
We agree about the postulate of possible completeness. The independent quantum states is "subjective".

Premise:

\begin{enumerate}
\item Quantum states provide a {\bf complete} description of physical properties.
\item 2.1 We need to postulate the existence of the observer. \\2.2 Every observer is not an element of physical property.
\end{enumerate}

\textbf{Quantum states is not a complete description of physical
reality.}\\

Observation: For instance, Fuchs' program \cite{Fuchs1} follow
this approach, we think with an important discrepancy between the
observer and the completeness. Qm is a complete description of
physical reality but the observer is not an element of physical
properties. The observer has not the same "substance". According
our assumptions the approach is the same subjective but quantum states
do not provide a "complete" description of physical reality (observer to remain not explained by QM)
According Fuchs instead we have a \textbf{complete} description of the physical reality and the
approach is not subjective, he affirm to be a realist, probably he admit the onticity of the observer.
It is possible to analyze in future others possibilities (next paper).

\section{Conclusion}

The conclusions are quite simple, according our philosophical assumptions is not possible an independent description of quantum states,
with all their implications, except third case, where the "a priori" seem possible but subjective.
We conclude that is not possible give an ontological proof of their existence, while is
always possible give a \textbf{subjective dependent existence}, utilizing the word
subjective in right way (observer presence)as defined in this paper, consequently the subject always \textbf{affect}
possible answers researched. Many different works affirm to have found an ontic or an epistemic $\mid\Psi>$, here we affirm,
under our philosophical conditions, that both are subjective.

\end{document}